\documentclass[aps,prl,twocolumn,superscriptaddress,nofootinbib,
nobalancelastpage,nobibnotes]{revtex4}
\usepackage{epsfig}

\begin{document}
\title
{Photon-photon interactions can be a source of CMB circular polarization}

\author{R. F. Sawyer}
\affiliation{Department of Physics, University of California at
Santa Barbara, Santa Barbara, California 93106}

\begin{abstract}

Photon-photon interactions, as described with the standard Heisenberg-Euler interaction, can transform plane polarization of the CMB into circular
polarization, in the period right after last scattering. We estimate the distribution of the resulting circular polarization parameters, as constrained by confining observations to very small angular regions of large plane polarization, and find 
results of the order of $10^{-9}$ for the Stokes parameter $V$ in some of these regions.
\end{abstract}
\maketitle

\section{1. Introduction}
Plane polarization of the cosmic background radiation (CMB) is induced by Compton scattering in an anisotropic photon bath \cite{bond} during the recombination period  \cite{peeb} \cite {zeld} at a temperature $\approx .25$ eV. But Compton scattering by itself cannot induce circular polarization. Therefore 
the potential for the detection of a tiny amount of circular polarization, should it exist, might provide evidence for some non-standard ingredients
in the early universe. For example it can result from anomalously large magnetic fields in the plasma prior to recombination \cite{gio},\cite{gio2}, or from the use of an exotic metric. It might result from photon-axion mixing effects during the course of propagation from the big-bang to the earth \cite{ag} or from beyond-the-standard model
effects \cite{bava}- \cite{al}. It might also result from the propagation through magnetized plasmas in galactic clusters \cite{silk}

Photon-photon interactions over a short period immediately after recombination provide an alternative mechanism by which circular polarization of photons
can be produced starting from plane polarization, even in the absence of magnetic fields or exotic physics. 
Beginning from the vacuum photon-photon interaction coming from the term of order $e^4$ in the one-loop effective
Lagrangian from QED (Heisenberg-Euler interaction)\cite{itz}, 
\begin{eqnarray}
L_{\rm H-E}=\int d^3 x{2 \alpha^2\over 45 m_e^4}[({\bf E^2-B^2)^2}+7 (\bf E\cdot B)^2]\,
\label{he}
\end{eqnarray}
where $m_e$ is the electron mass, $\alpha$ is the fine structure constant and we use units $\hbar=c=k_B=1$,
Kotkin and Serbo \cite{ks} showed that a polarized ``test" photon transiting a polarized laser beam can undergo
changes of its own polarization, through the interaction (\ref{he}), at a rate that is many, many of orders of magnitude faster than the rate of photon-photon scattering based on (\ref{he}) (see also \cite{rfs1}).
This stems from the polarization-dependent index of refraction induced by the laser beam in its interior. 
Coherence of the light, in the laser sense, is not required to obtain these effects.
The coherence that is exploited is that of completely forward interactions, that is, with
no momentum transfer between the ``test" photon and those of the surrounding bath.  The ``test"
photon picks up its polarization-change amplitude little by little from encounters with countless beam photons, and the changes
add up coherently.
In the application of this paper this takes place in a distance of order 50 megaparsecs, which is why what one might have guessed
to be an negligible effect can be appreciable. 

In the early universe we have nearly thermal ensembles of photons,
but with anisotropy that comes from the density perturbations, and angle-dependent linear polarization that arises from Compton scattering
within the anisotropic distributions. We focus here on the period immediately after the last photon scattering (i.e. at and after ``recombination"), and exclusively on the
development of some circular polarization.

\section{2. Outline of the calculation}

We follow the standard parameterization \cite{bond}
of the polarization density matrix $P$ for the ``beam photon" that we propose to observe. We take this photon to move with
momentum $q$ in the $\hat z$ direction, with basis vectors for polarization given by $(\vec \xi_1 =\hat x, \vec  \xi_2=\hat y)$
and define the Stokes parameters $Q,U,V,$ by,
\begin{eqnarray}
P={1\over 2} [\sigma_0+Q\sigma_3+ U \sigma_1+V \sigma_2]\,,
\label{pol1}
\end{eqnarray}
where $\sigma_i$ are Pauli matrices in the space with
basis $\vec \xi_1,\vec \xi_2$, augmented with the identity matrix $\sigma_0$.

Now we want to consider the interactions of the polarization of this photon with those of the entire assemblage
of photons in the region in which it was born, which we refer to as ``the cloud".
A photon in the cloud, moving in the $\theta,\phi$ direction, is taken to have the density matrix in polarization space, 
\begin{eqnarray}
P'={1\over 2} [b_0(\theta,\phi) \tau_0+q(\theta,\phi) \tau_3+ u(\theta, \phi) \tau_1+v(\theta,\phi) \tau_2]\,,
\label{pol2}
\end{eqnarray}
where the $\tau$'s are Pauli matrices acting on the polarization states of the cloud photon, in a basis now chosen as,
\begin{eqnarray}
\vec \eta_1=[\hat x \cos\phi+\hat y \sin \phi]\cos \theta-\hat z \sin \theta,
\nonumber\\
\vec \eta_2=-\hat x \sin \phi+\hat y \cos \phi\,.
\label{etabas}
\end{eqnarray}
There are four steps to what follows:

a). Taking all 16 matrix elements of $L_{\rm H-E}$ of (\ref{he}) between initial photon polarizations $\xi,\eta$ and final polarizations 
$\xi', \eta'$ for the
dead-forward kinematics  $\gamma_{\bf q_1}+\gamma_{\bf q_2} \rightarrow \gamma_{\bf q_1}+\gamma_{\bf q_2}$; then expressing
the result as a bilinear form in the matrices, $\sigma_i,\tau_i$. Since for our kinematics no photon changes its energy,
the resulting form serves as the complete effective Hamiltonian, $\mathcal{H}^{\rm eff}$, for the transformation process.
b). Determining the initial polarization density matrices, $P$ for the beam photon and
$P'(\theta,\phi)$ for the cloud photons. These are set by the final Compton scatterings of an anisotropic
photon distribution with a prescribed set of modes indexed by wavenumber $\hat k$, with intensities $d_k$.
 c). Integrating the evolution equation, beginning at the last scattering time, to get $V$, the circular polarization parameter for the beam photon, taking into account both the progressive weakening of
$\mathcal{H}^{\rm eff}$ from expansion of the universe, and the environmental change for the beam photon due to the finite wavelength of the acoustic perturbations of the cloud. 
d). Getting from standard sources the variances of the elements from which $V$ was constructed in c) ; then constructing the variance of $V$.

\section{3. Evolution equation.}
From direct evaluation of the matrix elements of (\ref{he}) for interaction of a beam photon with a single cloud photon we obtain the polarization dependent part of the effective Hamiltonian,
\begin{eqnarray}
\mathcal{H}^{\rm eff}_{q_1,q_2}=[{\rm vol.}]^{-1} W \sum_{i,j=0}^{i,j=3}  h_{i,j}(q_1, q_2,\theta, \phi)\, \sigma_i \tau_j \,,
\label{ham0}
\end{eqnarray}
where
\begin{eqnarray}
W={ 2 \alpha^2 \omega_1 \omega_2  \over 9 m_e^4}\,.
\label{w}
\end{eqnarray}
The non-vanishing elements of $h_{i,j}$ in (\ref{ham0}) are,
\begin{eqnarray}
h_{3,3}=-h_{1,1}=\cos (2 \phi) (1-\cos \theta)^2\,,
\nonumber\\
h_{3,1}=h_{1,3}= \sin(2 \phi) (1-\cos \theta)^2 \,,
\nonumber\\
h_{2,2}=(1-\cos \theta)^2 \,.
\label{heqns}
\end{eqnarray}

To get the approximate $\mathcal{H}^{\rm eff}$ for the interaction of the beam photon with the cloud, we integrate over a thermal distribution for the cloud, $|\vec q_2|$,
but keep a multiplicative angular dependence. The factor
of ${\rm vol.^{-1}}$ in (\ref{ham0}) is then replaced by the photon density, $n_\gamma$. Defining $\langle \tau_j \rangle $ as the cloud expectation values of the operators introduced in (\ref{pol2}), 
$\langle\tau_3 \rangle =q({\theta,\phi})$, $\langle \tau_1 \rangle =u(\theta,\phi)$,
we obtain from (\ref{ham0}) an $^{\rm eff}$ for the polarization evolution of the beam,
\begin{eqnarray}
\mathcal{H}^{\rm eff}(t)=w(t)\sum_{i=0}^3 \sigma_i c_i\,,
\label{ham2}
\end{eqnarray}
where,
\begin{eqnarray}
c_i= \int {d \Omega} \sum_{j=0}^3 h_{i,j}(\theta, \phi)\, \langle \tau_j \rangle \,,
\label{ham1}
\end{eqnarray} \,
and,
\begin{eqnarray}
 w(t)=2.1 \times 10^{-4}\Bigr ( {\omega_1\over \omega_1^*}\Bigr ) \Bigr({a(t_*)\over a(t)}\Bigr) \Gamma_C(t) \,.
\label{w}
\end{eqnarray}
Here $a(t)$ is the scale factor, $t_*$ the time at last scattering, and
$\Gamma_C(t)=8 n_e(t) \pi \alpha^2 m_e^{-2}/3$ is what the Compton scattering rate would have been, had there always been total ionization of the medium, i.e. no recombination. In calculating the numerical coefficient in (\ref{w}), we used $n_\gamma/n_e= 2 \times10^{ 9}$. $\Gamma_C$ serves here only as a convenient unit of rate. The factor 
$a(t_*)/a(t)$ in (\ref{w}) gives the effect of the red-shift on a cloud photon's energy after last scattering. The thermal average of $\omega_2$ at last scattering
has been absorbed in the coefficient in (\ref{w}). We normalized $\omega_1$ with its thermal average at last scattering, $\omega_1^*$, since this will be the same ratio
as will be observed on earth.

Once the factor $\langle   \tau_j \rangle (\theta,\phi)$ is known, and the angular integrals performed, the effective interaction $\mathcal{H}^{\rm eff}$ acts in the polarization space of the beam photon
as $\mathcal{H}^{\rm eff}=(c_1 \sigma_1+c_3\sigma_3)$ and the Heisenberg equation, ${d P/ dt}=-i [P,\mathcal{H}^{\rm eff}]$ for evolution of the circular polarization $V$ is simply, 
\begin{eqnarray}
{1\over 2}{d V \over dt}=(c_1 Q-c_3 U)w(t)
\label{dvdt}
\end{eqnarray}

\section{4. Initial conditions.}
Turning to the calculation of $c_1,c_3$ we first need the dependence of the cloud polarization on photon direction. Beginning
from the perturbation of the photon distribution in a direction $\hat n'$, with angular coordinates $\Omega'$,
 produced by an assemblage of sound waves in directions $\{\hat k\}$, and then calculating the polarization
density matrix, $\rho_{i,j}$ subsequent to a Compton scattering, we express the result in terms of the $\eta$ basis of (\ref{etabas}), 
obtaining,

\begin{eqnarray}
&\rho_{i,j}(\Omega)= \int d\Omega' \, \delta f(\Omega ')
\nonumber\\
&\times\Bigr [\vec \eta_i(\Omega)\cdot \sum_{m=1,2} \vec \eta_m(\Omega')\,\, \vec \eta_m(\Omega')\cdot \vec \eta_j(\Omega)\Bigr ]\,,
\end{eqnarray}
where  $\delta f(\Omega ')$
describes quadrupole photon density perturbations with wave numbers $\{k\}$. We take,
\begin{eqnarray}
\delta f( \Omega')=\sum _{\{k\}}{15 \over 16}\ d_k (\hat n \cdot \hat k)^2 \,,
\label{fk}
\end{eqnarray}
where we have dropped some spherically symmetric contributions of the modes, and normalization is arbitrary. Then $\mathcal{H}^{\rm eff}(t)$ of (\ref{ham2}) is constructed using,
\begin{eqnarray}
\langle \tau_3 (\theta,\phi) \rangle=(\rho_{1,1}-\rho_{2,2})/2
\nonumber\\
={1 \over 2} \sum_k d_k\sin^2 \theta_k{(1+\cos ^2\theta\,)}
\cos[2(\phi-\phi_k)] \,,
\label{tau3}
\end{eqnarray}
and
\begin{eqnarray}
\langle \tau_1(\theta,\phi) \rangle=(\rho_{1,2}+\rho_{2,1})/2
\nonumber\\
=  \sum_k d_k\sin^2 \theta_k \cos \theta
\sin [2(\phi-\phi_k)] \,.
\label{tau1}
\end{eqnarray}
In (\ref{tau3}) and (\ref{tau1}) we set $\theta=0,\phi=0$ to determine $Q$ and $U$, as induced by the perturbation (\ref{fk}),
\begin{eqnarray}
Q=\sum_k d_{k} \sin^2 \theta_k\cos 2\phi_k \,,
\nonumber\\
U=-\sum_k d_{k} \sin^2 \theta_k \sin 2\phi_k
 \,.
\label{QU}
\end{eqnarray}
From (\ref{tau1}), (\ref{tau3})and (\ref{ham1}) we have,
\begin{eqnarray}
c_1(t_*)=  {9  w(t_*)\over 8} U \,, \,c_3(t_*)= {9 \pi  w(t_*)\over 8} Q \,,
\label{c1c3}
\end{eqnarray}
for every configuration $\{k\}$.
Here $t_*$ is the time at recombination, defined as the temperature at which the ``visibility function" peaks. Now substituting 
 (\ref{QU}) and (\ref{c1c3}) into (\ref{dvdt}) we find $dV/dt=0$, at the time, $t_*$ of last scattering.
 But the relations (\ref{c1c3}) hold only at the recombination time, and will change during subsequent free-streaming.
 
 \section{5. Integration}
  $Q$ and $U$ for the beam photons are nearly frozen in time after the last scattering, as are the polarizations of individual photons in the cloud as well.
But $c_1(t)$ and $c_3(t)$, relating to the polarizations of the cloud evaluated at the position of the beam photon, change over a short
time region after $t_*$ as the photon passes through the cloud.\footnote{The cloud is time dependent on its own, but that matters less  since the acoustic wave velocity $< 3^{-1/2} c$.}  Thus, although the rate of $V$ generation vanishes at recombination (idealized as instantaneous) and is small immediately
afterwards, over a period of time the parameters $c_1(t), c_3(t)$ become disconnected from their values at $t=t_*$, as the beam photons move through the cloud. Clearly the perturbations of the cloud, evaluated at the position of the beam, which are of larger $k$ will change over a shorter time scale
than those of smaller $k$. This allows some $V$ to develop in the time period after $t_*$.

To make a rough estimate of the amount of $V$ that can be accrued, we begin with two questions: a) How much time is available before the expansion of the universe reduces the rates to the point of irrelevance? b) What is the minimum accoustical mode wave number $k$ over which $\{q,u\}$ can decouple from $\{Q,U\}$ soon enough to produce significant $V$?

Beginning with question a), we choose the orientation of the $x,y$ axes for a particular configuration of fluctuations such that $Q\ne 0, U=0$ for the beam. We also shift from time to scale factor, $\xi=a/a_*$, as the integration variable, where $a_*\approx .001$ is the scale factor at recombination.
From (\ref{dvdt})we obtain,
\begin{eqnarray}
V=2 Q  \int_{t^*}^\infty dt \, w(t) c_1(t)
\nonumber\\
\approx  {226\, Q  \,  w(t_*)\over \Gamma_{\rm C}(t^*)} \Bigr[ \int_{1}^{\xi_1}+\int_{\xi_1}^{\infty} \Bigr] d\xi\, \xi^{-7/2} c_1(\xi) \,.
\label{V1}
\end{eqnarray}
where we used $dt=d\xi (\xi H)^{-1}$ and $H$ is the Hubble rate at time $t$. In (\ref{w}) we substituted $\Gamma_C\approx 113 (\xi)^{-3/2} H$ (ref. \cite{dod} eq. 3.46).

Note that $c_1$ is not damped by pure expansion, since it is generically the ratio of a density to a density. But it is constrained by the requirement that $c_1(1)=0$, by virtue of (\ref{c1c3}) with $U=0$. Our crude approach will be to take $c_1(\xi<\xi_1)=0$ in an initial time period and $c_1(\xi>\xi_1)={\rm const.}$; in the latter region unconstrained by a connection to the beam polarization. 

To address the value of $c_1$ in the second region, we turn to question b), beginning from the fact that in the free-streaming era the generic time dependence of a mode $(k, \ell)$ is
as $j_\ell[k (\eta-\eta_*)]$ where $\eta$ is the conformal time and $\eta_*$ is its value at recombination. 
The peak contribution from an acoustic mode of wave-number $k$ to the present, $ \eta=\eta_0$, sky comes when $\ell=k(\eta_0-\eta_*)$, with $\eta_0\approx 32 \eta_*$. This gives $k\approx \ell/(32 \eta_*)$.

Choosing instead a small scale factor change after $t_*$ corresponding to $\xi-1=\delta \xi$, we use $\delta \eta/\eta_*=(1/2) \delta \xi$  and again look for the peak contribution from  $j_\ell[k (\eta-\eta_*)]$. Phase randomization at $k\,  \delta\eta\approx 1$ coming from the argument of $j_\ell$ then sets in at a value $\ell_{\rm min}\approx 64/ (\xi_1 -1)$. We now implement our two-region approach by setting the first integral to zero, and in the second integral including only the fluctuations with $\ell> \ell_0$.
We have checked this method of estimation by replacing the division of scale $\xi$  into two regions with a subdivision into six regions, each with different $\ell_{\rm min}$'s as determined by the above construction. The results are roughly the same as we obtain for the two region case provided that for the latter case we choose $\xi_1=1.25$  for the division point in (\ref{V1}), giving $\ell_{\rm min} \approx 256$.  
\section{6. Distribution functions}
The program from here on is to determine the variance of $c_1(\xi )$, as generated by the random fluctuations in the distributions of polarization densities at $t_*$, to be determined by polarization measurements for $\ell >256$ on the present sky, or theoretically from assumed primordial fluctuation spectra. The standard deviation of $V$ then is calculated  using (\ref{c1c3}), (\ref{w}) and (retaining only the second $\xi$ integral) eq. (\ref{V1}),
where, $ \langle u u \rangle_{\ell>256}$ is the variance of the polarization at one angle as determined in the usual way from the partial wave variances indexed with $\ell$ except for the deletion of modes with $\ell<256$. 
\begin{eqnarray}
\langle VV\rangle^{1/2} \approx  .047  (Q^2+U^2)^{1/2}
\nonumber\\
\times[\langle u u \rangle_{\ell>256}+ \langle q q \rangle_{\ell>256}]^{1/2}\,.
\label{ans1}
\end{eqnarray}
Here we are now adopting a fixed choice for the $x,y$ axis orientation, rather than choosing it such that $U=0$. The factor $(Q^2+U^2)^{1/2}$ is the beam's plane polarization which is to be directly observed; and the final factor, which remains to be computed, is the standard deviation of the cloud's polarization.

As input data we shall use the present
theoretical \cite{zal} plus observational \cite{page} results for $\langle q_\ell^2 \rangle +\langle u_\ell^2 \rangle=2\langle u_\ell^2 \rangle$, rather than going back to the basic power spectra derived from inflationary models. The literature quotes results on these polarization fluctuations in terms of effective temperature changes
$\Delta T_\ell$. However, in (\ref{ans1}), they are defined as (polarization number density)/(background number density). The conversion mechanics
is simply to replace $\,^\circ \mu K \,$ in all results by $(3/2.78) \times 10^{-6}$ 

From the WMAP results we then find 
\begin{eqnarray}
\langle q_ \ell q_\ell \rangle+\langle u_ \ell u_\ell \rangle\approx \times .06 \times 10^{-6} \pi \ell^{-1}\,,
\label{ans3}
\end{eqnarray}
after smoothing out oscillations, between $\ell =100$ and $\ell=1000$ or so, after which $\langle q_ \ell q_\ell \rangle$ levels off and then starts to decrease with increasing $\ell$.

Inverting the expansion in spherical harmonics we have,
\begin{eqnarray}
&\langle[ q (\theta,\phi) q(\theta',\phi')+u (\theta,\phi) u (\theta',\phi')]\rangle =
\nonumber\\
&\sum_{\ell,m}\sum_{\ell',m'}\Big \langle \Bigr [q^{(\ell,m)} q^{(\ell',m')}+u^{(\ell,m)} u^{(\ell',m')}\Bigr ]\Big \rangle 
\nonumber\\
&\times  Y_{\ell,m}(\theta ,\phi)Y_{\ell',-m'}(\theta' ,\phi') \,.
\label{t3t3}
\end{eqnarray}
The correlation function in (\ref{t3t3}) vanishes except when $(\ell,m)=(\ell 'm')$. Taking the limit in which $\theta,\phi, \theta',
\phi'$ all approach zero we get,
\begin{eqnarray}
&\langle [qq+uu]\rangle_{\ell >256} =\sum_{\ell>256,|m|\le \ell} \langle [q_\ell q_\ell+ u_\ell u_\ell ] \rangle
\nonumber\\
&\times Y_{\ell,m}(0,0) Y_{\ell,-m}(0,0) \approx 4.0 \times 10^{-10}\,,
\label{QQ}
\end{eqnarray}
where in performing the sum in the last line we took, 
\begin{eqnarray}
 \langle [q_\ell q_\ell+ u_\ell u_\ell ] \rangle 
\approx .07\pi\ell^{-1} \exp(-\ell/1000) \times10^{-12},
\nonumber\\
\,
\end{eqnarray}
now in dimensionless units.
in accord with the above remarks.
From (\ref{QQ}) and (\ref{ans1}) we then obtain the standard deviation of $V$, $\sigma_V$

 \begin{eqnarray}
 \sigma_V= .11 \times 10^{-5} (Q^2+U^2)^{1/2} 
 \end{eqnarray} 
 
 Then if a cut is made to select only those regions in which $(Q^2+U^2)^{1/2} >2 \times 10^{-5}$, which should be plentiful, one time in twenty we would find $V>5 \times 10^{-10}$ Given this standard deviation we would expect to find $V>.6 \times10^{-10}$ around $5\%$ of the time. The degree of polarization also can
 also be enhanced by looking only at the highest band of photon energies, in view of the proportionality of the fundamental rate
 (\ref{w}) to $\omega_1$ (which earlier had been replaced by a thermal average). This could get us into the range of $V\approx2 \times 10^{-9} $.

 \section{7. Discussion}

There are distinctive features of the $V$ signal that is produced by the mechanism.  The proportionality to photon energy is one that can be tested, in principle,
in measurements on a single spot in the sky. This could provide a way of distinguishing the mechanism from others, e.g. one based on linear into circular transformation in a magnetized plasma living inside a galactic cluster along the transit path \cite{silk},
or one in which the magnetized plasma is taken to exist before and at the time of recombination \cite{gio}.

We have not attempted calculation of the effects of our mechanism in the case of tensor perturbations. In the presence of photon-photon interactions they may be more efficient at producing $V$ than are scalar perturbations, for cases in which amplitudes are comparable. Indeed if 
the numbers reported in the recently published Bicep2 results \cite{bicep} are sustained, tensor perturbations could be the dominant contributor to circular polarization.

Our use of the basic interaction (\ref{he}) was perturbative, in the sense that the quantity $V$ remains very small, and had insignificant back-reaction at later times through the evolution equations on the quantities $Q,U$ that were specified in the initial condition. 
In some other astrophysical venues
in which basic interaction (\ref{he}) can be expected to play a role, for example in the core of $\gamma$-ray bursts, the back reactions will matter. 
The effective interaction for polarization exchange will still be given by (\ref{ham0}).
In this connection, we note the close correspondence of the formalism used here with that used in predicting effects of the neutral current
$\nu-\nu$ interaction on the neutrino flavor evolution near the core of the supernova \cite{sr}. In the latter problem there is a large literature devoted to sudden flavor transformations due to the non-linear equations \cite{lit}. We might expect the same with respect to photon polarizations in, e.g.,
gamma ray burst analysis.
 
To summarize: we have calculated some ingredients for a respectable calculation of the variance of $V$. Up through eq. (\ref{c1c3}) our results are authoritative, but the estimates that follow are very crude.  The numbers perhaps justify a more complete approach. While our results are orders of
magnitude smaller than presently planned observations could detect \cite{limit}, we hope that they might encourage the development of observational
technology that could make detection possible.

This work was supported 
in part by NSF grant PHY-0455918.


\begin{thebibliography}{99} 
\bibitem{bond} J. R. Bond, G. Efstathiou, MNRAS {\bf 226}, 655 (1987)
\bibitem{peeb}P. J. E. Peebles, Astrophys. J.,{\bf 153}, 1 (1968)
\bibitem{zeld}Ya. B. Zel'dovich, V. G. Kurt, R. A. Sunyaev, Soviet Physics JETP, Vol. 28, p.146
\bibitem{gio}M. Giovannini, Phys.Rev. {\bf D80}, 123013 (2009); arXiv:0909.3629
\bibitem{gio2} M. Giovannini, K. E. Kunze, Phys.Rev. {\bf D78}, 023010 (2008)
\bibitem{ag} N. Agarwal, P. Jain, D. W. McKay, J- P. Ralston,
	Phys. Rev. {\bf D78}, 085028 (2008); arXiv:0807.4587
\bibitem{bava} E. Bavarsad, M. Haghighat, Z. 
Rezaei, R. Mohammadi, I. Motie, M. Zarei, Phys. Rev. {\bf D81}, 084035 (2010);
arXiv:0912.2993
\bibitem{fi} F.Finelli, M. Galaverni, Phys.Rev. {\bf D79}, 063002 (2009); arXiv:0802.4210
\bibitem{al}S. Alexander, J. Ochoa , A. Kosowsky,
Phys.Rev. {\bf D79}, 063524, (2009); arXiv:0810.2355
\bibitem{silk}A. Cooray , A. Melchiorri, J. Silk, Phys.Lett. {\bf B554}, 1 (2003); arXiv:astro-ph/0205214
\bibitem{itz}e.g., C. Itzykson and J-B. Zuber, ``Quantum Field Theory", McGraw-Hill (New York, 1980), Eq. (4-125 )
\bibitem{ks}  G. L. Kotkin and V. G. Serbo, Phys. Lett. {\bf B413}, 122 (1997)
\bibitem{rfs1}R. F. Sawyer, Phys. Rev. Lett. {\bf 93}, 133601 (2004); arXiv:hep-ph/0404247
\bibitem{dod} Scott Dodelson, ``Modern Cosmology", Academic Press (New York, 2003)
\bibitem{zal} M. Zaldarriaga , U. Seljak, Phys.Rev.{\bf D55},1830 (1997); arXiv:astro-ph/9609170
\bibitem{page} L. Page, et al., Astrophys. J. Suppl. 170:335,2007;  arXiv:astro-ph/0603450
\bibitem{bicep}The BICEP2 Collaboration, Phys. Rev. Lett. {\bf 112}, 241101 ( 2014); arXiv:1403.3985 
\bibitem{sr} G. Sigl and G. Raffelt, Nuc. Phys. {\bf B406}, 423 (1993)
\bibitem{lit} e.g.,G. Raffelt, S. Sarikas, D. de Sousa Seixas, Phys. Rev. Lett. {\bf 111}, 091101 (2013), and references contained therein.
\bibitem{limit}R. Mainini, et al., JCAP{\bf 08},033 (2013); arXiv:1307.6090
\end{thebibliography}
\end{document}